\documentstyle[12pt,epsf]{article}
\newcommand{\mfrac}[2]{\mbox{$\frac{#1}{#2}$}}

\newcommand{\bd}[1]{{\bf #1}}

\newtheorem{theorem}{Theorem}

\newtheorem{definition}{Definition}

\newcommand{\lphi}{\mbox{$\displaystyle\lim_{\phi\rightarrow\infty}$}}
\newcommand{\wbar}{\mbox{$\overline{W}_V$ }}
\newcommand{\fbar}{\mbox{$\overline{f'}$ }}

\makeatletter                    
\@addtoreset{equation}{section}  
\makeatother                     

\begin{document}
\title{Scalar Field Cosmologies and the Initial Space-Time Singularity}
       
\author{{\bf Scott Foster}}
\date{Department of Physics and Mathematical Physics\\
University of Adelaide, Adelaide, Australia
\\}

\maketitle

\begin{abstract}
The singularity structure of cosmological models whose matter content consists of a scalar field with arbitrary non-negative potential is discussed. The special case of  spatially flat FRW space-time  is analysed in detail using a dynamical systems approach which may readily be generalised to more complicated space-times.  It is shown that for a very large and natural class of models a simple and regular past asymptotic structure exists. More specifically, there exists a family of solutions which is in continuous 1-1 correspondence with the exactly integrable massless scalar field cosmologies, this correspondence being realised by a unique asymptotic approximation. The set of solutions which do not fall into this class has measure zero. The significance of this result to the cosmological initial value problem is briefly discussed.    
\end{abstract}

\section{Introduction}
The global structure of solutions of Einstein's equations, and in particular, the apparent existence of a cosmological space-time singularity has long been of  interest to cosmologists. Considerable  research effort during the 1960s and early 70s culminated in the detailed description of the cosmological singularity  given by Belinskii, Khalatnikov and Lifschitz (BKL) 
\cite{bellip1,belip1.5,belip2}  and in the well known singularity theorems of Hawking, Penrose {\it et al} \cite{hawkandell}. The latter provide rigorous proof of the existence of singularities under very general circumstances, and have long been regarded as demonstrating that singularities are inevitable in realistic cosmologies.

 In the absence of a complete theory of quantum gravity, the classical singularity remains the only unambiguous means of describing the boundary of classical space-time and is therefore of importance in the formulation of a well posed initial value theory for cosmology. The cosmological initial conditions play a particularly prominent role in inflationary cosmology, which aims to provide a mechanism whereby the universe evolves from an arbitrary initial state  toward the highly symmetric, isotropic universe we observe today \cite{guth}. At present, the question of whether or not inflation sucessfully achieves these goals remains unclear  \cite{goldpir,gibbetal}, largely, it is suggested, due to the difficulty in assigning meaningfull initial conditions in the absence of a global description of solution space.  

 The global description of inflationary theories is complicated by the fact that a number of inflationary scenarios, such as  chaotic inflation \cite{linde},  place the onset of inflation  close to 
the boundary of the Plank era. On the other hand, inflation itself seems to arise from an essentialy classical mechanism and can be modelled by the evolution of a scalar field in a classical space-time
background. Both the classical and quantum mechanical limiting cases are therefore of interest. It must be noted that the characteristic pre-inflationary time-scale for scalar field models will depend on the scalar field self-interaction term as well as some measure on the set of initial conditions, and there is no {\it a priori} reason why it should be of the order of the Plank time.    

The structure of the classical singularity requires careful re-examination in the inflationary context since the strong energy condition \cite{hawkandell}, which plays an important role in both the singularity theorems and the construction of the BKL singularities, can not be strictly applied in the early universe if  inflation occurs \cite{barrowdef}.  

 In fact, it can easily be demonstrated by explicit example that there exist scalar field cosmologies  which possess no singularity at all (see for example \cite{mad2}). Other well known solutions (ie, the power law  inflation solutions ) possess singularities but no particle horizons \cite{lucmat} and it is  possible, by prudent choice of the potential, to engineer exact Friedmann Robertson Walker (FRW) solutions with scale factor $a(t)$ exhibiting almost any type of behavior one desires.  
 
In the present paper, we shall consider scalar field models generated
by an action functional of the following form:
\begin{equation}\label{eq:action}
S[g_{\mu\nu}, \phi ]=\int d^4x\sqrt{-g}(R + \mfrac{1}{2} g^{\mu\nu}\partial_\mu\phi\partial_\nu\phi+V(\phi).)
\end{equation}
where $R$ is the Ricci scalar, $g$ is the determinate of the metric
 $g_{\mu\nu}$
and        
$V$ is an arbitrary non-negative potential. (Note that we shall always work with units $c=8\pi G= 1$ and adopt the metric signature $(- + + +)$.)
 
Although  in what follows  $g_{\mu\nu}$ is implicitely taken to be the physical  metric on space-time, it should be kept in mind that effective scalar field theories of the form (\ref{eq:action}) arise naturally, via conformal transformations, in the
context of generalised scalar tensor theories and higher order theories of gravity 
\cite{whitt,barcot,hall1,hall2,dic,sin1,har,ger,alb}.
 For such theories $g_{\mu\nu}$ will be equivalent to the physical metric up to a conformal factor determined by $\phi$.

  Quantitative calculations of global behavior of the solutions generated by (\ref{eq:action}) requires a detailed knowledge of the global form of the scalar field
 potential, a  task that  is fundamentally difficult since the energy density of the scalar field can be expected to diverge during the approach to a singularity. Interactions modeled by a simple power law, for example, would seem unlikely to remain valid over such a range. Furthermore, there are no known physical fields which have emerged as strong candidates for the ``inflaton'' field present in the early universe. It is by no means clear whether the scalar field driving inflation should be viewed as a fundamental physical field at all,
or merely a convenient heuristic model which elucidates certain (qualitative) dynamical features of the early universe. The close relationship between scalar fields and conformal transformations, coupled with the likely significance of higher order gravitational lagrangians at high energies, is suggestive of the latter point of view.
  
 In accordance with the above remarks it shall be our approach to place strong emphasis on the structural features of scalar fields as a class of models, rather than the quantitative details of specific physical scenarios for inflation. We shall approach the problem from a dynamical point of view. If one insists that the energy density of the scalar field  be non-negative everywhere,
 then it can be shown that the level hypersurfaces of the scalar field are space-like 
and (almost) always foliate space-time \cite{thesis}. Therefore, a global 
cosmological time arises naturally and one can consider the evolution of the 
induced 
3-metric and scalar field, together with appropriately defined time 
derivatives (or conjugate momenta),  as an infinite dimensional dynamical system. 

The easiest way to make further progress is by making use of the long 
wavelength
 approximation in which one neglects terms of second order in the spatial 
derivatives, thereby de-coupling the spatial and temporal dependence of the 
dynamical variables . The 
evolution at
 each spatial point is then governed by a finite set of ordinary differential
 equations equivalent to the evolution equations for spatially homogeneous
  space-times. The underlying physical assumption is that kinetic terms (ie,
 expansion and shear) diverge faster than spatial gradients and therefore 
dominate at very early times. For a more detailed discussion of long wavelength approximation schemes in inflation and  cosmology in general see, for example, \cite{com1,com2,salbond,sal,eard}. 

 The long wavelength dynamics are characterised by the spatially homogeneous scalar field 
cosmologies, the solutions of which reside in a 7-dimensional phase space.
 It turns out, however, that many of the interesting asymptotic features 
characteristic of scalar field cosmologies  are already present in the 
 relatively simple spatially flat Friedmann Robertson Walker  
space-time models which, in addition to being an important subset of
 spatially homogeneous metrics in its own right, has the advantage that 
it
 can be represented by a two dimensional vector field. This allows one to 
visualise the flow and thereby gain considerable insight into the geometric
 structure of solution space, before tackling the higher dimensional problem.

In particular, a number of authors have made fruitful investigations of inflationary structure in various FRW scalar field models  using phase plane techniques (\cite{bel1,burdbarr,hall2}) and, importantly, this approach has proven to be readily generalisable to more complex space-times (eg, \cite{burdbarr}). Accordingly, the  remainder of the paper shall be devoted to a detailed
 analysis of the general past asymptotic structure of spatially flat FRW scalar
 field
 cosmologies from the geometric point of view. It is claimed that the 
techniques
 developed below can readily be generalised to more general space-times as is discussed in greater detail in  \cite{thesis}.   

 
In Sections 2-4  we  write down the field equations, assuming a general non-negative potential $V$, as a two dimensional dynamical system and investigate some of the elementary features of the flow. It turns out that the scalar field almost always diverges asymptotically in the past making it necessary to introduce some additional regularity conditions on $V$ (which is unbounded in general) that allow us to compactify phase space and analyse the structure of the system at infinity, without the need for an explicit choice of potential. In Section 5 it is shown that under very general conditions there always exists a generic class of solutions with an initial singularity which possesses a simple and elegant structure. These solutions are in continuous 1-1 correspondence with the exactly integrable massless scalar field cosmologies ($V\equiv 0$), which 
provide a unique asymptotic approximation. In Section 6 these results are used to prove a global singularity theorem for FRW scalar field cosmologies and the cosmological significance is briefly discussed in Section 7.  

Throughout the text major results are summarised as theorems. Reference is made to a number of results from dynamical systems theory. Readers unfamiliar with dynamical systems techniques, and, in particular, the properties of center manifolds are referred to the excellent introductory texts by Guckenheimer and Holmes \cite{gucholms} and Wiggins \cite{wigg} which include comprehensive lists of major references. 
  
\section{The Field Equations.}\label{flatrw:prel}
Consider the FRW line element
\begin{equation}\label{eq:rw}
 ds^2 =- dt^2 + a(t)^2\sum_{i=1}^3 ( dx^i)^2.
\end{equation}   
We define the expansion $K(t)$ as the function
\begin{equation}\label{eq:adot}
K=3\frac{\dot{a}}{a},
\end{equation}
where the ``dot'' indicates differentiation with respect to time $t$. K may be interpreted physically as the rate of expansion of the spatial volume element.

The evolution equations obtained from variation of (\ref{eq:action}) may be expressed as
a dynamical system in the variables $(K, \phi, \dot{\phi})$  as follows \cite{goldpir}:
\begin{eqnarray}
\frac{d\phi}{dt}&=&\dot{\phi}\nonumber\\
\frac{d\dot{\phi}}{dt}& =  & -K \dot{\phi}- V'(\phi)\label{eq:S1}\\
\frac{dK}{dt}& = & -\mfrac{3}{2}\dot{\phi}^2 \nonumber
\end{eqnarray}
subject to the constraint
\begin{equation}\label{algcon}
K^2 =  3V(\phi) + \frac{3}{2}\dot{\phi}^2.
\end{equation}
The constraint equation actually makes (\ref{eq:S1}c) redundant since $K$ can
 be expressed as a function of $\phi$ and $\dot{\phi}$. 

It can easily be verified that, for non-negative potentials, the set $K=0$ is invariant under (\ref{eq:S1}) with constraint (\ref{algcon}). Since no two orbits may  intersect it follows that the sign of $K$ is invarient. 
The physical expanding solutions of  (\ref{eq:S1}) therefore consist of 
 trajectories
 in the 2-dimensional phase space
\[
\Omega =\{(K,\phi, \dot{\phi}):K > 0,K^2=3V(\phi) + \mfrac{3}{2}
                                 \dot{\phi}^2\}\subset
 R^3.
\]
It shall be convenient to  replace $K$ and $\dot{\phi}$ with  the new set of coordinates
\begin{equation}\label{var1.5}
x=\frac{1}{K}\hspace{1cm}y=\sqrt{\mfrac{3}{2}}\frac{\dot{\phi}}{K}
\end{equation}
and introduce a new time coordinate
 \begin{equation}
\label{taudef}
\tau=\ln v(t) + \tau_0
\end{equation}
where $v=a^3$ is the spatial volume element and $\tau_0$ is some constant.
 $\tau$ is well defined since $v$ is strictly increasing. Furthermore, $v$ 
goes to zero if and only if $K$ goes to infinity \cite{hawkandell} so (\ref{taudef}) implies that $K\rightarrow \infty$ as $\tau\rightarrow 
-\infty$. Differentiating (\ref{taudef}) we find (recalling that $K=\dot{v}/v$)
 
\begin{equation}
\frac{d}{dt}=K\frac{d}{d\tau}.\label{eq:tau}
\end{equation}
In terms of these new coordinates the dynamical equations for the system
 become
\begin{eqnarray}
\frac{dx}{d\tau}&=&y^2x \nonumber\\
\frac{dy}{d\tau}&=&-y-\sqrt{\mfrac{3}{2}}x^2V'({\phi})+y^3 \label{eq:S4}\\
\frac{d{\phi}}{d\tau}&=&\sqrt{\mfrac{2}{3}}y \nonumber
\end{eqnarray}
subject to the constraint
\begin{equation}\label{eq:con4}
y^2+3x^2V({\phi})=1.
\end{equation}  
If we imagine tracing a trajectory of the above system backwards in time then it may be seen from the above equations that both $x$ and $y$ remain bounded.
However, it is not possible to eliminate $\phi$ from the above equations since the constraint (\ref{eq:con4}) is degenerate at $x=0$ and $V=0$. The analysis of the above system under the condition that $\phi$ remained bounded would be relatively straitforward, particularly since within some bounded region it would be appropriate to model $V$ by a polynomial of sufficiently high order, greatly simplifying the problem. If $\phi$ diverges, on the other hand, then any such approximation schemes will almost certainly be  inappropriate. It is therefore of considerable interest that $\phi$ almost always diverges as shall now be shown.

\begin{theorem}\label{th:unbound}
Assume that $V$ is $C^3$.
 Let $p$ be a point in $\Omega$ and
let $O^-(p)$ be
the past orbit of $p$ under (\ref{eq:S4}).  
Then $\phi$ is almost always unbounded on $O^-(p)$.
\end{theorem}
Proof:

Let $p\in\Omega$ be such that $\phi $ is bounded on $O^-(p)$.
It is therefore clear that  $O^-(p)$
 is contained in a compact subset of (the closure of) $\Omega$.
 It is intuitively obvious, and indeed follows from a
 basic result of dynamical systems theory (See \cite{wigg}, Proposition 1.1.14)
 that the trajectory must asymptotically approach some limit set $\alpha(p)$.  It is easy to convince oneself that $\alpha (p)$ must itself be a union of orbits. Since $x$ is monotonic, it must clearly be constant on $\alpha(p)$. (\ref{eq:S4}a) therefore implies that either $y=0$ or $x=0$. The latter case would imply, by (\ref{eq:con4}), that $y^2 =1$ on $\alpha(p)$ in which case it would follow from (\ref{eq:S4}c) that $\phi$ is unbounded, in contradiction to our previous assertion. Therefore, $\alpha(p)$ lies on the plane $y=0$. 

Inspection of (\ref{eq:S4}) reveals that the
only possible invariant sets lying on the line $y=0$ are equilibrium points 
$(x_0, 0, \phi_0)$, where $V'(x_0)=0$ and $x_0^2= \frac{1}{3}V(\phi_0) > 0$.

 In order to complete the proof we need only to demonstrate that no such 
equilibrium point is a past asymptote to an open set of trajectories on 
$\Omega$. Note that the equilibrium points of (\ref{eq:S4}) correspond to 
the equilibrium points of (\ref{eq:S1}), which may be expressed explicitely
 as a 2-dimensional (unconstrained) autonomous system by making the 
natural projection onto 
the plane
\begin{equation}
X=\phi, \hspace{1cm}Y=\dot{\phi}.
\end{equation}
 and eliminating $K$, using (\ref{algcon}), to obtain
\begin{equation}
\begin{array}{rcl}
\dot{X}&=&Y\\
\dot{Y}&=&-V'(X)-\left (3V(X)+\frac{3}{2}Y^2\right )^{\frac{1}{2}}Y.
\end{array}
\label{eq:S2}
\end{equation}      
Let $ q=(X_0,Y_0)$, be an equilibrium point of ({\ref{eq:S2}) satisfying
\begin{equation}
Y_0=0\hspace{1cm} V'(X_0)=0\hspace{1cm}V(X_0)=V_0>0
\end{equation}

  We can classify the stability of equilibrium points by computing the
eigenvalues of the matrix of derivatives of the vector field (\ref{eq:S1}) at
 $q$ . We find
\begin{equation}
J_q=\left(\begin{array}{cc}
     0        & 1\\
     -V''(X_0)& -\sqrt{3V_0}
      \end{array}\right)    
\end{equation}
The eigenvalues of $J_q$ are:
\begin{equation}
\lambda=\frac{1}{2}\left(-\sqrt{3V_0}\pm\sqrt{3V_0- 4V''(X_0)}\right)
\label{eq:eig}\end{equation}
At least one of these eigenvalues must always be negative. It follows from
 the Center Manifold Theorem (See \cite{gucholms}, Theorem 3.2.1) that there exists a local stable manifold 
intersecting $q$. That is, a 1 or 2-dimensional invarient submanifold 
of $\Omega$ on which all solutions close to $q$ exponentially approach $q$ as 
$t\rightarrow\infty$. The existance of a stable manifold of dimension $n > 0$
 implies that all solutions past asymtotic to $q$ must lie on an unstable
 manifold or center manifold of dimension $2-n < 2$\footnote[2] {If the second eigenvalue,
 above,
 is non-zero, then q is a hyperbolic fixed point and the result follows from the Hartman-Grobman linearisation Theorem (\cite{gucholms}, Theorem 1.3.1 )
If there
 exists a zero eigenvalue, then the topology of the flow near $q$ is non-trivial and is characterised by a unique local center manifold which acts as an exponential attractor in the forward time sense and therefore a repeller in the reverse time sense (See \cite{vand}, Corollary 4.7). Briefly a center manifold is the non-linear analogue of the center subspace from the theory linear ODEs. It can be loosely defined as an invariant manifold to the flow which intersects $q$ and is tangent to the zero eigenvector at $q$.}. This completes the proof. 
$\Box$
\\ 
  
It is of interest to dwell briefly on the physical meaning of the  equilibrium points, above. Firstly observe that they correspond to
 maximally symmetric solutions of
Einstein's equations. With $V_0>0$, $q$ represents a vacuum 
de Sitter solution with scale factor
\[
a(t)= e^{\sqrt{\frac{V_0}{3}} t}.
\]

The sign of the eigenvalues depends on the sign of $V''(X_0)$. If $V''(X_0)$ 
is positive ($X_0$ a  local minimum of $V$) or negative 
($X_0$ a local maximum),
 then $q$ will be a stable sink or a saddle respectively.
 In the latter case exactly two solutions are past asymptotic to $q$.
  It is interesting to note that these solutions correspond
 physically to unstable de Sitter solutions, similar to those discussed
 by Barrow 
\cite{barrowdef}. They asymptotically approach de Sitter 
space in the past
 and possess neither singularities or particle horizons. 
 
If  $V''(X_0)=0$, one of the eigenvalues is zero and the behavior near $q$
 will be characterized by a local center manifold $(X, Y(X))$, tangent to the
 $X$-axis  at $q$, whose dynamics will be dependent on $V$.

\section{Conditions on the Potential}
 Theorem \ref{th:unbound} makes it clear that
 in order to investigate the generic past
 asymptotic behavior of the system (\ref{eq:S1}) we need look at  the behavior  near $\phi=\pm\infty$. 
 In order to do this we shall find it necessary to make some further restrictions on the behavior of $V$ as $\phi$ diverges. 

Firstly we shall restrict attention to potentials which are non-zero for large $\phi$. In particular, this excludes the case where $V$ has compact support. 
This will not lead to a significant loss of generality however, since  equations (\ref{eq:S1}) with $V$ identically zero are exactly integrable and the solution is well known (we shall write it down explicitly later on). 

 Non-zero potentials may be classified qualitatively according to their degree of ``flatness'' which we choose to  characterise by the ratio $V'/V$, where $V'$ is the derivative of $V$.  Intuitively we expect that the flatter potentials are more likely to exhibit inflation, for example. Indeed, the condition that  $V'/V$ be small plays an important roll in the slow roll approximation and is often stated as an explicit condition for inflation \cite{barpar}. 

Exponential type behavior would be characterised by $V'/V$ tending to a non-zero constant (as $\phi\rightarrow\pm\infty$) whilst $V'/V$ tending to zero would indicate slower than exponential behavior, such as a polynomial. Extremely steep potentials, or those with oscillatory type behavior, would be without a well defined limit for $V'/V$ as $\phi\rightarrow\infty$. 

In the remainder of this section we focus on the behavior of $V$ near $\phi =+\infty$, an analogous discussion applies near $\phi =-\infty$:
\begin{definition}\label{def:subexp}
Let $V:R\rightarrow R$ be a $C^2$ non-negative potential. Let there exist 
some $\phi_0>0$ for which $V(\phi)> 0$  
 for all $\phi>\phi_0$ and some number $N$ such that the function 
$W_V:[\phi_0,\infty)\rightarrow R$,
\[W_V(\phi)=\frac{V'(\phi)}{V(\phi)} - N \]
 satisfies 
\begin{equation}
\lim_{\phi\rightarrow\infty}W_V(\phi)=0.\label{eq:Wlim}
\end{equation}
Then we say that $V$ is Well Behaved at Infinity (WBI) of exponential order $N$.
\end{definition}
It is important to point out that $N$ may be 0, or even negative. Indeed the class of WBI functions of order 0 is of particular interest, containing all non-negative polynomials and other 'flattish' functions.

A trivial consequence of the above definition is the following:
\begin{theorem}\label{th:Vlim}
let $V$ be WBI 
 of exponential order $N$ then, for all $\lambda>N$,
\begin{equation}
\lim_{\phi\rightarrow\infty}e^{-\lambda\phi}V(\phi)=0.
\end{equation}
\end{theorem}

We shall now define a procedure for classifying the smoothness of WBI 
functions at infinity. We simply say that 
 a   WBI function is of class $k$ if there
exists a sufficiently smooth coordinate transformation $z=f(\phi)$, which takes infinity to the origin, for which $W_V$ is
 mapped to a $C^k$ function in some 
neighbourhood of the origin. Furthermore, we should only admit coordinate 
transformations 
 $f$ which are useful from the point of view of dynamics, in that they induce
 a smooth map on the tangent bundle 
(ie smooth vector fields are mapped to smooth vector fields).
 Therefore  the first derivative of $f$ with
respect to $\phi$, that is $f'$, should also be expressible as a $C^k$ function of
 $z$.   

We
 take note of the following convention,
 which will be assumed henceforth. If we have some coordinate 
transformation $z=f(\phi)$ which maps a neighbourhood of infinity to a 
neighbourhood of the origin, then if $g$ is a function of $\phi$,
 $\overline{g}$ is the function of $z$ whose domain is the range of
$f$ plus the origin, which takes the values;
\[
\overline{g}(z)=\left\{\begin{array}{rcr} g(f^{-1}(z))&,&z>0\\
                                     \lphi g(\phi)&,&z=0 \end{array}\right.
\] 
          
\begin{definition}\label{def:sexpk}
A $C^k$ potential $V$ is class k WBI if it is
 WBI and if  there exists $\phi_0>0$ and a coordinate
 transformation $z=f(\phi)$ which maps the interval $[\phi_0,\infty)$ onto
$(0, \epsilon]$, where $\epsilon=f(\phi_0)$ and $\lphi f=0$, with 
 the following additional properties: 
\begin{tabbing}
i)\hspace{0.4cm}\=  $f$ is $C^{k+1}$ and strictly decreasing.\\
ii)            \>the functions $\wbar(z)$ and $\fbar(z)$ are $C^k$ on 
the closed interval $[0,\epsilon]$.\\
iii)           \> ${\displaystyle \frac{d\overline{W}_V}{dz}(0)=\frac{d\overline{f'}}{dz}(0)=0.}$
\end{tabbing} 
\end{definition}
We designate the set of all  class k WBI functions ${\cal E}^k_+$ .
The third requirement in the above definition, that the first derivatives
of \wbar and \fbar vanish at the origin, is an additional condition on {\em the transformation} $f$ which has been imposed for reasons of convenience that will become clear later. It will not 
effect the class of a potential since if the first derivatives of {$\fbar$} exist
 at the origin (for some coordinate transformation $z$)
 they can always be 
made zero by choosing the new coordinate $z'=z^{\frac{1}{2}}$.

 As illustrated by 
Table~I, a diverse range of qualitative behavior is incorporated within the framework of ${\cal E}^k_+$ potentials. It can easily be verified that all non-negative polynomials and linear combinations of exponentials are
 ${\cal E}^\infty_+$. The very general inflationary potentials discussed by Barrow \cite{barbar} and Barrow and Parsons \cite{barpar} are also ${\cal E}^\infty_+$. Furthermore, sufficiently ``smooth''  distortions to the shape of the potential should not change the situation very much. For example, it can be shown that any potential for which   $W_V$ is $C^2$  will be ${\cal E}^2_+$, provided that  the second derivative of $W_V$ vanishes faster than $\phi^{-(2+\epsilon)}$ as $\phi\rightarrow\infty$, for some $\epsilon > 0$
\cite{thesis}\footnote[2]{Conversely, if $W_V''\geq O(\phi^{-2})$ then 
$W_V\geq O(\ln\phi)\rightarrow\infty$ as $\phi\rightarrow\infty$.}. This is a sufficient but not a  necessary condition, as can be readily verified for the quasi-exponential potential
\[
V(\phi)=e^{\frac{\phi}{\ln{\phi}}}
\] 
using the transformation $z=1/(\ln{\phi})^{\frac{1}{2}}$.

\begin{table}\label{tab:subexp}
\begin{tabular}{l|l|l|l|l}
 $V(\phi)$&$W_V(\phi)$ & $z=f(\phi)$&$\wbar (z)$ &$\fbar (z)$ \\ \hline
$\left|\frac{\lambda}{n}\right|\phi^{n}$&$n\phi^{-1}$
 &$\phi^{-\frac{1}{2}}$
 &$n z^2$&
$-\frac{1}{2}z^3$\\[5pt]
$e^{\lambda\phi}$&0&$\phi^{-1}$&0&$-z^2$\\[3pt]
$2e^{\lambda\sqrt{\phi}}$ &$\lambda\phi^{-\frac{1}{2}}$
&$\phi^{-\frac{1}{4}}$ &$\lambda z^2$&$-\frac{1}{4}z^5$\\[3pt]
 $\ln{\phi}$&$(\phi\ln\phi)^{-1}$&$(\ln\phi)^{-1}$&$ze^{-\frac{1}{z}}$&$-ze^{-\frac{2}{z}}$\\[3pt]
$\phi^2\ln{\phi}$&$2\phi^{-1} +(\phi\ln\phi)^{-1}$&
$(\ln\phi)^{-1}$&$(2+z)e^{-\frac{1}{z}}$&$-ze^{-\frac{2}{z}}$\\[3pt]
\end{tabular}
\caption{Simple examples of WBI behavior at large $\phi$. $n$ and $\lambda$ are arbitrary constants.}
\end{table}

 As a final note before proceeding, observe
 that the condition ${\displaystyle  \frac{d\fbar}{dz}(0)=0} $ in Definition
\ref{def:sexpk} may be expressed equivalently (using the chain rule and
inverse function theorem )  as
\[
\lim_{\phi\rightarrow\infty}\frac{f''}{f'} =0.
\]
It follows that $1/f'$ is WBI of exponential order $0$ and hence by Theorem~\ref{th:Vlim}
that it is  exponential dominated.
 This implies in turn that $1/f$ is also exponential dominated. The
function $f$ must therefore obey the following condition: 
 for all $\epsilon >0$. 
\begin{equation}\label{fexplim}
\lim_{\phi\rightarrow\infty}\frac{e^{\epsilon\phi}}{f'} =\lim_{\phi\rightarrow\infty}\frac{e^{\epsilon\phi}}{f}=0.
\end{equation}

\section{The Flow Near $\phi=\infty$}
In this section we shall characterise the qualitative structure of the flow 
near $\phi=\infty$. As in the previous section, analogous results hold near
 $\phi =-\infty$. We shall assume that $V\in{\cal E}^2_+$.
 Define
the set $\Omega_\epsilon\subset \Omega$ to be the set of points in $\Omega$ 
for which $\phi > \epsilon^{-1}$ where $\epsilon$ is any positive  constant.
  $\epsilon$ is chosen sufficiently small so as to avoid any points where 
$V=0$, thereby ensuring that $W_V$ is well defined.  
 We now  make the coordinate transformation 
\begin{equation}\label{eq:coords}
 (x, y,\phi ) \mapsto (x, y, z),\hspace{1cm}z=f(\phi )
\end{equation}
on $\Omega_\epsilon$, where $f(\phi)$ tends to 0 as 
$\phi\rightarrow\infty$ and has been chosen so that conditions i)-iii) of Definition
~\ref{def:sexpk} are sastisfied, with $k=2$.   

Substituting these new coordinates into (\ref{eq:S4}) and making use of the constraint equation (\ref{eq:con4}) to eliminate $x$ we obtain the 2-D dynamical system

\begin{eqnarray}
\frac{dy}{d\tau}&=& -(y + \mfrac{1}{\sqrt{6}}(\overline{W_V}(z)+N))(1-y^2) \nonumber \\
\frac{dz}{d\tau}&=&\sqrt{\mfrac{2}{3}}\overline{f'}(z)y \label{eq:S7}
\end{eqnarray}
We may identify $\Omega_\epsilon$ with its projection into $R^2$  so that we have:
\[ \Omega_\epsilon =\{0<z<f(\epsilon^{-1}),
-1<y<1\} \]
 The variable $x$ can now be treated as a
function on $\Omega_\epsilon$ defined by the constraint equation which becomes.
\begin{equation}\label{eq:constraint}
y^2+3x^2\overline{V}(z)=1.
\end{equation} 
 The directional derivative of $x$ along the
flow generated by (\ref{eq:S7}) may be obtained directly from Equation (\ref{eq:S4}a).

Since $\fbar$ and $\wbar$ are $C^2$ at $z=0$ we may extend (\ref{eq:S7}) onto the boundry of $\Omega_\epsilon$ to obtain a $C^2$ system on the
 closure of $\Omega_\epsilon$, that is  $\tilde{\Omega}_\epsilon$.

Recall also, from Definition \ref{def:sexpk}, that $\fbar$ and $\wbar$ vanish at the origin and are each second order or higher in $z$, and that $\fbar $ is negative on $\Omega_\epsilon$.

As a simple example, consider the special case of the power law potential
\[ V(\phi)=\frac{\lambda}{n}\phi^n.\]
where $\lambda$ and $n$ are positive constants. Introducing the 
coordinate $z=\phi^{-\frac{1}{2}}$ and substituting the expressions
 $\wbar$ and $\fbar$ given in
 Table~1 into equation~(\ref{eq:S7}) gives the system
\begin{eqnarray}
\frac{dy}{d\tau}&=&-y+ y^3-\frac{n}{\sqrt{6}}z^2(1-y^2) \nonumber \\
\frac{dz}{d\tau}&=&-\frac{1}{\sqrt{6}}z^3y. \label{eq:powerpot}
\end{eqnarray}
Observe that it is perfectly regular everywhere on $\tilde{\Omega}_\epsilon$
(in fact, everywhere on $R^2$).
}
\\

\begin{figure}[htb]
\epsfysize=6cm
\epsfxsize=12cm
\begin{center}
\leavevmode
\epsfbox{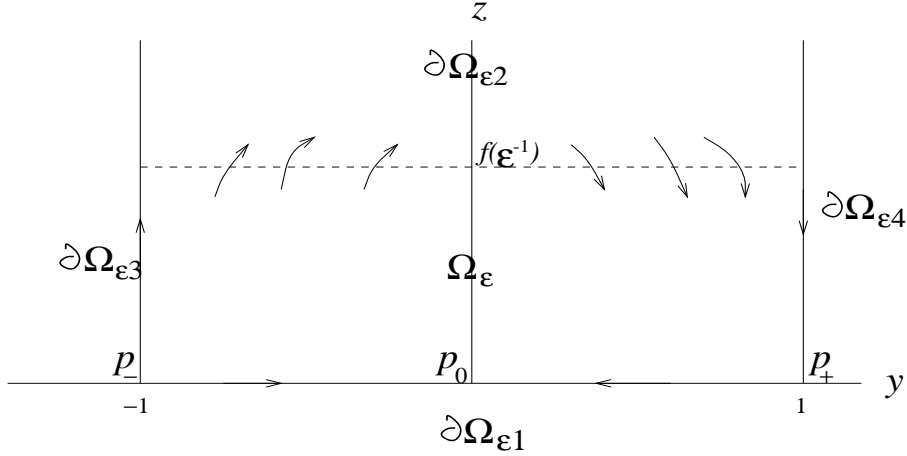}
\end{center}
\caption{Schematic representation of the compact set $\tilde{\Omega}_\epsilon$ showing
fixed points (with N=0) and the sense of orbits on the boundary.}
\label{fig:ohmhat}\end{figure}

Returning to the general case, a schematic representation of  $\tilde{\Omega}_\epsilon$ is shown in Fig. \ref{fig:ohmhat}. 
The boundary $\partial{\Omega}_\epsilon$
is the union of the unphysical sets $\partial{\Omega}_{\epsilon 1}$,
$\partial{\Omega}_{\epsilon 3}$ and
 $\partial{\Omega}_{\epsilon 4}$, as well as the physical set 
$\partial {\Omega}_{\epsilon 2}$ on which the observables
$K$, $\phi$ and $\dot{\phi}$ are finite. $\partial{\Omega}_{\epsilon 1}$,
$\partial{\Omega}_{\epsilon 3}$ and $\partial{\Omega}_{\epsilon 4}$
are each invariant manifolds of (\ref{eq:S7}).
 $\partial{\Omega}_{\epsilon 2}$ is the only element of the boundary 
which is transverse to the flow. It is through this set that physical 
trajectories enter and leave $\tilde{\Omega}_\epsilon$.

 By  (\ref{eq:S7}b) it can be seen that the sign of
$\frac{dz}{d\tau}$ is opposite to the sign of $y$. 
 Thus trajectories may enter $\tilde{\Omega}_\epsilon$ only on the
right half of the plane and leave only on the left half of the plane.

 The points
 $( -\mfrac{1}{\sqrt{6}}N, 0)$,
 $(1,0)$ and $(-1, 0)$ are the only  equilibrium 
points, and shall be referred to as $p_N$, $p_+$ and $p_-$ respectively. Note that the equilibrium point $p_N$ only occurs for potenials of exponential order $|N| < \sqrt{6}$. In what follows we shall always assume this to be the case. It turns out that this bifurcation represents a transition between inflationary models and non-inflating models as is well known from the structure of exponential potential models (eg,  \cite{burdbarr}).

The structure of the flow  (\ref{eq:S7}) close to the equilibrium points $p_\pm$ can most readily be deduced from the total derivative of the vector field $(\frac{dy}{d\tau}, \frac{dz}{d\tau})$
evaluated at the respective equilibrium points:
\begin{equation}
J(p_\pm)=\left (\begin{array}{cc} 2(1-\frac{N}{\sqrt{6}})&0\\
                                 0&0 \end{array}\right ).
\end{equation}
The eigenvalues of $J(p_\pm)$ are $2(1-\frac{N}{\sqrt{6}})$ and 0, with corresponding eigenvectors $\hat{e}_y=(1,0)$ and $\hat{e}_z=(0,1)$ respectively.

It therefore follows from Fig. \ref{fig:ohmhat} that: $\partial{\Omega}_{\epsilon 1}$ contains local  unstable manifolds for $p_+$ and $p_-$ (ie, solutons close to $p_\pm$ having $z=0$ are exponentially repelled from the respective equilibrium point);  $\partial{\Omega}_{\epsilon 3}$ is a center manifold of $p_+$ and;
  $\partial{\Omega}_{\epsilon 3}$ is a center manifold of $p_-$. Recall that by center manifold, we just mean an invariant manifold tangent to the zero-eigenvector. Because the linear part of (\ref{eq:S7}) vanishes, the restricion of the flow to such sets is slow compared to the exponential decay/growth tangent to the $y$-axis.

Consequently, all solutions in the neighbourhood
of $p_+$ or $p_-$ will rapidly approach
the respective center manifolds as they evolve backwards in time, and the stability of $p_+$ and $p_-$ depends on the asymptotic behavior
of the solutions on $\partial{\Omega}_{\epsilon 3}$ and $\partial{\Omega}_{\epsilon 4}$ respectively (\cite{wigg},  Theorem 2.12). 

\begin{figure}[htb]
\epsfysize=4cm
\epsfxsize=10cm
\begin{center}
\leavevmode
\epsfbox{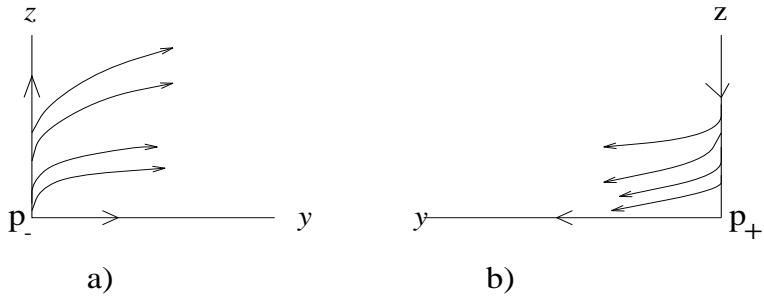}
\end{center}
\caption{The stability of the equilibrium points $p_+$ and $p_-$.}
\label{fig:source}\end{figure}

We may  write down the solution of (\ref{eq:S7}) on $\partial{\Omega}_{\epsilon 3}$ and
$\partial{\Omega}_{\epsilon 4}$ by solving the
 differential equation
\begin{equation}
 \frac{dz}{d\tau}=\pm\overline{f'}(z)\sqrt{\mfrac{2}{3}}.
\end{equation}
Recalling that $z=f(\phi)$ this is just 
\[ \frac{d\phi}{d\tau}=\pm\sqrt{\mfrac{2}{3}}\]
which yields
\begin{equation}\label{phi22}
\phi=\sqrt{\mfrac{2}{3}}(\pm\tau+ \tilde{\phi})
\end{equation}
where $\tilde{\phi}=\phi(0)$ is a positive constant. Thus 
\begin{equation}
z=f\left(\sqrt{\mfrac{2}{3}}(\tilde{\phi}\pm\tau)\right).\label{eq:zas}
\end{equation}
The flow is directed towards the y-axis on 
$\partial\Omega_{\epsilon 4}$ and away from the y-axis on
$\partial\Omega_{\epsilon 3}$ as shown in Fig. \ref{fig:ohmhat}.
Thus, 
$p_+$ is a saddle (Fig.~\ref{fig:source}a) and $p_-$ is a source. 
(Fig.~\ref{fig:source}b).

We now turn our attention to the remaining equilibrium point $p_N= (-\frac{N}{\sqrt{6}}, 0 )$.
Evaluating the derivative of the vector field at this point gives
\begin{equation}
J(p_0)=\left (\begin{array}{cc} -1 + \frac{N^2}{6}&0\\
                                 0&0 \end{array}\right ).
\end{equation}
The eigenvalues are $-1 + \frac{N^2}{6}$ and 0. This implies that there locally exists
a 1-dimensional center manifold $C_N$ through $p_N$, which is tangent to the $z$-axis
at $p_N$ (and, in general, only at $p_N$), and a 1-dimensional stable manifold tangent to (and indeed a subset of)
 $\partial{\Omega}_{\epsilon 1}$. In accordance with our previous comments on the properties of center manifolds it is clear that $C_N$ is an exponential attractor on a sufficiently small neighbourhood of $p_N$ and it is intuitively obvious from the geometry, and indeed is rigorously true (\cite{vand}, Corollary 4.7), that  any solutions past asymptotic to $p_N$ must lie on 
the center manifold; ie:
\begin{theorem}\label{th:pN}
Assume $V\in{\cal E}^2_+$ with $N^2<6$. All solutions asymptotic in the past to $p_N$ lie on a unique orbit $C_N$ 
 (corresponding to a single physical cosmology) which is (locally) a center manifold 
of $p_N$.   
\end{theorem}
For our purposes it is sufficient to observe that Theorem~\ref{th:pN} implies that at most one cosmology can originate at $p_N$ and, therefore, that $p_-$ is the only generic source in $\partial{\Omega}_{\epsilon}$. It turns out however, that the center manifold $C_N$ is of fundamental physical interest in its own right, being intimately related to  the inflationary dynamics of a particular model. $C_N$ corrosponds to the Inflationary Attractor emphasised by Liddle {\em et al} \cite{lidetal}, and plays a role directly analogous to that played by de Sitter space in cosmologies with positive cosmological constant. As such its significance goes well beyond the FRW ``toy cosmologies'' being discussed in this paper, since it acts as an attractor in the full space of Einstein's equations, and allows inflation to be characterised precisely in a very general context.  

Furthermore, the unique cosmology living on $C_N$ accounts for virtually all 
known exact spatially flat FRW solutions (see \cite{barpar}, Table 1), including all singularity free and 
horizon free solutions (excluding the trivial de Sitter cosmologies). 
The most significant exception is, of course, the  
general solution for $V=0$ (massless scalar field) which shall be discussed 
in further detail below.  

The fundamental dynamical role of $C_N$ in scalar field cosmologies, and its relationship to inflation, shall be explored in more detail in a forthcoming paper.

\begin{figure}[htb]
\epsfysize=5cm
\epsfxsize=10cm
\begin{center}
\leavevmode
\epsfbox{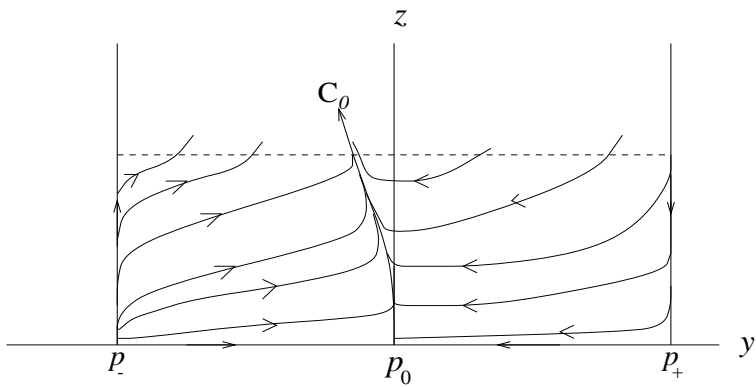}
\end{center}
\caption{Schematic representation of the flow on $\Omega_{\epsilon}$ for $\wbar >0$.}
\label{fig:glob1}\end{figure}

 Before proceeding to the next section, we may summarise the results above by  constructing a phase portrait:  Fig. \ref{fig:glob1}  sketches the flow  for the special case, $N=0$, $\overline{W_V}(z)>0$, illustrating the qualitative behavior and in particular
the topological structure of the system near $\phi=\infty$. 

\section{The Space-Time Singularity}

In this section we investigate the behavior close to  the generic source $p_-$ in more detail, and demostrate that it corresponds to a space-time singularity.
The past  asymptotic behavior of solutions close to $p_-$ will be approximated
by the exact solution on the center manifold  up to an exponential error term
 (this may be readily verified directly). Thus for $\tau$ sufficiently large and 
negative we may write
\begin{equation}
y=-1 + O(e^{\gamma \tau})\hspace{1cm}z=f\left(\sqrt{\mfrac{2}{3}}(\tilde{\phi}-\tau)\right)
 +O(e^{\gamma \tau})
\label{eq:assoln} 
\end{equation}
For some positive constant $\gamma$. Note that evaluating the inverse 
$f^{-1}(z)$ in the above expression and using the mean value theorem 
and the expressions (\ref{fexplim}) we find that 
\begin{equation}\label{eq:assoln2}
\phi=\sqrt{\mfrac{2}{3}}(\tau+ \tilde{\phi}) + O(e^{\beta \tau})
\end{equation}
 for some positive $\beta$, as we would expect from (\ref{phi22}).
 
 Let us now express the asymptotic solution (\ref{eq:assoln}) wholly in
terms of our observables. As $y\rightarrow - 1$,
equation (\ref{eq:S4}a) becomes
\begin{equation}
\frac{dx}{d\tau}=x(1 +O(e^{\gamma \tau})).
\end{equation}
Integrating gives the first order expression
\begin{equation}
x=x_0e^\tau.\label{eq:orig1}
\end{equation}
By (\ref{eq:tau})
\begin{eqnarray}
t&=& \int x d\tau \nonumber\\
 &=& x_0e^\tau + t_i\label{eq:orig2}
\end{eqnarray}
which implies that $t\rightarrow t_i$ as $\tau\rightarrow -\infty$.
Substituting (\ref{eq:orig2}) into (\ref{eq:orig1}) and using
(\ref{var1.5}) gives
\begin{equation}
K=(t-t_i)^{-1}. \label{eq:Has}
\end{equation}
Note that $K\rightarrow \infty$ as $t\rightarrow t_i$, thus the source
$p_-$  corresponds to an initial space-time singularity.
Substituting (\ref{eq:orig2}) and (\ref{eq:Has}) into (\ref{eq:assoln}) (recalling the definition of $y$) we find
\begin{equation} 
\phi=-\sqrt{\mfrac{2}{3}}\ln{\frac{(t-t_i)}{c}}\hspace{1cm}\dot{\phi}=
-\sqrt{\mfrac{2}{3}}(t-t_i)^{-1} \label{eq:phias}
\end{equation}
where $c= x_0e^{\tilde{\phi}}$ is a positive constant. 

 The asymptotic solution (\ref{eq:Has}), 
(\ref{eq:phias}) is the exact solution of the system (\ref{eq:S1})
when $V$ is identically zero; ie the massless scalar field. Thus there
exists a generic class of cosmologies which, in
 a sufficiently small neighbourhood of the singularity,
behave, approximately, as though the matter content were a massless 
scalar field.

In order to make the above statement more precise it is necessary to do
two things. Firstly, we shall obtain
an error estimate for equations ~(\ref{eq:Has}) and
 (\ref{eq:phias}) close to $t=t_i$. Secondly, we shall demonstrate
that close to $t=t_i$ there is a continuous 1-1 correspondence between
massless scalar field cosmologies and cosmologies of the model $V(\phi)$
which are past asymptotic to $p_-$. 
 We shall assume, without loss of generality, that $t_i=0$.  
\begin{theorem}\label{th:approx}
Let $V\in {\cal E}_+^2$ have  $N^2 < 6$. 
There exists a neighbourhood ${\cal N}(p_-)$ of $p_-$ such that for all $p\in {\cal N}(p_-)$
 the trajectory $\psi_p$ approaches  $p_-$ asymptotically in the past and the solution may be written:
\begin{eqnarray} 
K&=&t^{-1}+O(\epsilon_V(t))\nonumber\\
\phi&=&-\sqrt{\mfrac{2}{3}}\ln\frac{t}{c}+O(t\epsilon_V(t))\label{eq:approx}\\
\dot{\phi}&=&-\sqrt{\mfrac{2}{3}}t^{-1} + O(\epsilon_V(t))\nonumber
\end{eqnarray}
where
\begin{eqnarray}
\epsilon_V(t)&=&tV(-\sqrt{\mfrac{2}{3}}\ln t ).\label{eq:error}
\end{eqnarray}
\end{theorem}
Note that from (\ref{eq:phias}), 
(\ref{eq:orig2})
 and Theorem 
~\ref{th:Vlim}
\begin{equation}\label{limV}
 \lim_{\tau\rightarrow -\infty}e^{\sqrt{\frac{3}{2}}\alpha \tau}\overline{V}(z(\tau))=
   \lim_{t\rightarrow 0}t^\alpha V(\phi(t)) = 0 \;\;\forall\;\alpha \geq 
\sqrt{\mfrac{2}{3}} N.
\end{equation}
In particular this confirms that for $N^2 <6$ the error terms above 
 are dominated by the first order terms and, furthermore, for $N^2 < \frac{3}{2}$, the error terms tend to zero (uniformly, as we shall see below) as $t\rightarrow 0$. 
  
Proof of Theorem:

We proceed by calculating a second order 
term for the function $x$. Any collection of higher
order terms which are to be discarded will be denoted by a generic $h$.
By higher order terms we mean, more precisely, that the ratio of $h$ to all
undiscarded terms tends to zero as $\tau\rightarrow -\infty$.
We have
\begin{equation} 
\frac{dx}{d\tau}= y^2x
\end{equation}
where we know that $y^2\rightarrow 1$ as $\tau\rightarrow
-\infty$ and $x$ approaches the first order solution (\ref{eq:orig1}).

From (\ref{eq:constraint})
\begin{eqnarray}
           y^2&=&1-3\overline{V}(z)x^2\nonumber\\
              &=&1-3x_0^2\overline{V}(z)e^{2\tau} + h.\label{eq:y2nd}
\end{eqnarray}
Thus, 
\begin{equation}\label{eq:lnxdot}
\frac{d\ln{x}}{d\tau}=1-3x_0^2\overline{V}(z)e^{2\tau} + h 
\end{equation}
We will now show that 
\begin{equation}\label{eq:limV2}
\int\overline{V}(z)e^{2\tau}d\tau = 
\frac{\overline{V}(z)e^{2\tau}}{\gamma} + h
\end{equation}
where $\gamma = 2 - \sqrt{\mfrac{2}{3}}N$. Define the function $T= V e^{-N\phi}$. Since $V$ has exponential order $N$ it follows from (\ref{eq:assoln2}) that 
 
\[
\int\overline{V}(z)e^{2\tau}d\tau = 
\int\overline{T}(z)e^{\gamma\tau}d\tau  + h.
\]
 Consider the 
indefinite integral
\[
I(\tau)=\int_{\tau_0}^{\tau} \overline{T}(z) e^{\gamma\tau}d\tau 
\]
 Integrating by
 parts gives,
\begin{eqnarray}
I(\tau)&=&\mfrac{1}{\gamma}\overline{T}(z)e^{\gamma\tau}|_{\tau_0}^{\tau}
        -\frac{\sqrt{2}}{\gamma\sqrt{3}}\int_{\tau_0}^{\tau}e^{\gamma\tau}\overline{T'}(z)yd\tau\nonumber\\
       &=&I_1(\tau)+I_2(\tau) \nonumber
\end{eqnarray}
Consider the second term. 
Let $\delta(\tau)=\sup_{\tau'<\tau}\frac{\sqrt{2}}{\gamma\sqrt{3}}|\wbar(z(\tau'))|$, 
then, recalling that $y^2< 1$;
\begin{eqnarray}
|I_2(\tau)|&=&\frac{\sqrt{2}}{\gamma\sqrt{3}}\left |\int_{\tau_0}^{\tau}e^{\gamma\tau}\overline{T}(z)\wbar(z)yd\tau\right |\nonumber\\
         &<&\delta(\tau)|I(\tau)|\nonumber\\
         &\leq&\delta(\tau)(|I_1(\tau)|+|I_2(\tau)|)\nonumber\\
         &<&\frac{\delta(\tau)}{1-\delta(\tau)}|I_1(\tau)|
\end{eqnarray}
for $\tau$ sufficiently small.
 Letting $\tau_0$ go to $-\infty$ demonstrates (\ref{eq:limV2}), as required.

We may now integrate (\ref{eq:lnxdot}), using  (\ref{eq:limV2}) and
 (\ref{limV})
  to obtain,  
\begin{eqnarray}
x&=& x_0(e^\tau - \mfrac{3}{\gamma}x_0^2\overline{V}(z)e^{3\tau}) +h\label{x}
\end{eqnarray}
Integrating again to obtain a second order expression for $t$ we get
\[
t=x_0(e^\tau -\mfrac{1}{\gamma}x_0^2\overline{V}(z)e^{3\tau}) + h 
\]
where the second term on the left hand side has been estimated in precisely
the same way as $I(\tau)$ above. This expression may be inverted, to second 
order, to give
\begin{equation}\label{eq:tausec}
x_0e^\tau = t + \frac{\overline{V}(z)t^3}{\gamma} + h.
\end{equation}
substituting this into (\ref{x}) and recalling that $x=1/K$ we find;
\begin{equation}\label{eq:H2nd}
K=t^{-1}+ \frac{2}{\gamma}V(\phi)t +h.
\end{equation}
In a similar manner we may obtain an error estimate for $\phi$. Recall that
\[
\frac{d\phi}{d\tau}=\sqrt{\mfrac{2}{3}}y
\]
which, may be integrated using (\ref{eq:y2nd}), to obtain
\begin{equation}\label{eq:phi2nd}
\phi(t)=-\sqrt{\mfrac{2}{3}}\left(\ln\frac{t}{c} - \mfrac{1}{2\gamma}V(\phi)t^2
\right) +h.
\end{equation}
where (\ref{eq:tausec}) was used to replace $\tau$ by $t$ in  the above expression. Similarly,
\begin{eqnarray}
\dot{\phi}&=&\sqrt{\mfrac{2}{3}}Ky\nonumber\\
          &=&-\sqrt{\mfrac{2}{3}}\left(t^{-1}  - \mfrac{1}{\gamma}V(\phi)t\right)
 + h.\label{eq:phidot2nd}
\end{eqnarray}  
Clearly, the second term on the right hand side of (\ref{eq:phi2nd})
tends to 0 as $t\rightarrow 0$. This allows us to Taylor expand $V$
about $\phi=-\sqrt{\frac{2}{3}}\ln\frac{t}{c}$  to obtain,
\begin{eqnarray}
V(\phi(t))  &=&V\left(-\sqrt{\mfrac{2}{3}}\ln\frac{t}{c}\right)\left (1+
               b W_V\left(-\sqrt{\mfrac{2}{3}}\ln{\frac{t}{c}}\right)V(\phi)t^2\right )+h\nonumber
\end{eqnarray}
where $b$ is a constant. Substituting this equation into (\ref{eq:H2nd}), 
 (\ref{eq:phi2nd}) and (\ref{eq:phidot2nd}) completes the proof. $\Box$ \\

\begin{theorem}\label{th:rw1-1}
Let $V\in {\cal E}^2_+$ be of exponential order $N^2 <6$. For sufficiently small neighbourhoods ${\cal N}(p_-)$ of $p_-$, the set of physical orbits intersecting ${\cal N}(p_-)$ is homeomorphic to the set of massless scalar field cosmologies; ie,  For each $c>0$ there exists a unique 
cosmology  satisfying (\ref{eq:approx}), and this corrospondence is continuous.
\end{theorem}
Proof:

Although the above theorem seems intuitively rather obvious, its proof is non-trivial due largely to the fact that precise details of the transformation between the trajectories of the system (\ref{eq:S7}) and the physical solutions with respect to proper time $t$ is not known.

   Firstly,
recall that for each solution past asymptotic to $p_-$ there exists 
\begin{equation}\label{cdet}
c=x_0e^{\tilde\phi}
\end{equation}
where $x_0$ and $\tilde\phi$ are as defined in (\ref{eq:orig1}) and
 (\ref{eq:assoln}). More precisely we may define
\begin{equation}\label{xandphidef}
x_0=\lim_{\tau\rightarrow -\infty}e^{-\tau}x(\tau)\hspace{1cm}\tilde{\phi}=
\lim_{\tau\rightarrow -\infty}
\left(\tau +  \sqrt{\mfrac{3}{2}}\phi(\tau)\right).
\end{equation}
 Each of these constants depend
 on the point $p$
  at which we take
$\tau= 0$ but  for a given solution $\psi_p(\tau)$
 a translation $\tau\mapsto \tau + 
\tau_0$ (which moves $p$ along its orbit ) 
leaves $c$ invariant. This reflects the fact that each orbit corresponds to
only one physical cosmology.

In order to see this more clearly we make use of the fact that
 $\phi$ diverges and  is strictly decreasing with $\tau$ close to $p_-$. It
 follows that we may represent 
 any orbit as a graph $(\phi, y(\phi))$. Substituting 
(\ref{xandphidef}) into (\ref{cdet}) and remembering that $x$ is a function
of state we have:
\begin{equation}
c=\lim_{\phi\rightarrow \infty}e^{\sqrt{\frac{3}{2}}\phi}x(\phi, y(\phi)).
\end{equation} 
Thus $c$ depends only on the graph $y(\phi)$ (i.e. the orbit) and is
independent of the initial point $p$.

Let ${\cal N}(p_-)$ be a small neighbourhood of $p_-$ and let $y(\theta; \phi)$
be a continuous one-parameter family of orbits covering ${\cal N}(p_-)$. $\theta$ is a bounded  non-negative real parameter with the property that  $\theta_2 > \theta_1$ implies $y(\theta_2;\phi) > y(\theta_1; \phi)$. Since two orbits may never cross, the above family of orbits will be well defined as long as ${\cal N}(p_-)$ is sufficiently small that each orbit can be writen as a graph wrt $\phi$. $\theta$ may be viewed as an angular coordinate.

We shall now demonstrate that $c$ is a continuous, 1-1 function of $\theta$.
This will immediately prove the first part of the theorem and
 the second part of the theorem (continuity) is then a trivial consequence of
 the continuity of the flow and the
coordinate transformations (\ref{taudef}), (\ref{eq:coords}).

Define the function 
$$
C(\theta;\phi) = x(\phi, y(\theta;\phi))e^{\sqrt{\frac{3}{2}}\phi}
$$
then $c$ is, by definition, the limit of $C$. Taking the derivative of this expression with respect to $\phi$ and making use of (\ref{eq:S4}a,b) we have 
\begin{equation}\label{Cdiff}
\frac{dC}{d\phi}= \sqrt{\frac{3}{2}}(y + 1)C
\end{equation} 
 Also, from (\ref{eq:constraint}) we may obtain an explicit expression for $C$
\begin{equation}\label{cexpl}
C(\theta; \phi)^2 = \frac{e^{\sqrt{6}\phi}}{3V(\phi)}(1-y(\theta;\phi)^2).
\end{equation}  
 It is immediately apparent from this expression that $c$ is a continuous
  since
the orbits $y(\theta;\phi)$ are uniformly continuous in $\theta$.     
Also, since $y$ is strictly increasing with $\theta$ (for fixed $\phi$),
(\ref{cexpl}) implies that $C(\theta;\phi)$ is strictly 
increasing with $\theta$. 

 Let $\theta_2>\theta_1$ and, for convenience, denote the graphs $y$ and $C$ along the respective
orbits by $y_1,C_1$ and $y_2,C_2$. Taking the derivative of the
difference $C_2(\phi)-C_1(\phi)$ using (\ref{Cdiff}) we obtain
\begin{equation}
\frac{d}{d\phi}\left(\ln\frac{C_2(\phi)}{C_1(\phi)}\right)=\sqrt{\frac{3}{2}} (y_2(\phi)-y_1(\phi))>0                            
\end{equation}
where we have used the fact that $y_2-y_1>0$.
 Integrating from some $\phi_0$ to $\infty$ we obtain 
\[
\ln\left(\frac{c(\theta_2)C_1(\phi_0)}{c(\theta_1)C_2(\phi_0)}\right) 
                >0.
\] 
from which it follows that 
\[
\frac{c(\theta_2)}{c(\theta_1)}>\frac{C_2(\phi_0)}{C_1(\phi_0)}>1
\]
since $C(\theta;\phi_0)$ is strictly increasing. Thus
$c$ is strictly increasing with $w$ and hence 1-1.

This proves the theorem. $\Box$ \\

Theorem \ref{th:rw1-1} demonstrates that $V$ is truly dynamically insignificant
 in the neighbourhood of the 
singularity $p_-$ in that the family of solutions which asymptotically
approach  $p_-$ are  completely
characterized by the solution space of the massless scalar field cosmological
model. We shall discuss some cosmological consequences of this result in Section 7 but before doing so we shall make use of the above results to prove a global singularity theorem for scalar field cosmologies.

\section{A Global Singularity Theorem}
 In
order to complete the analysis of the global system 
it is, of course, necessary to also look at the
behavior near $\phi=-\infty$. This can be done easily by
observing that the equations (\ref{eq:S1}) are invariant under the 
transformation 
\begin{equation}\label{mirror}
(\phi ,\dot{\phi})\longmapsto -(\phi ,\dot{\phi}) \hspace{1.3cm} V\longmapsto U
\end{equation}
where $U(\phi)=V(-\phi)$. Thus, for a particular potential $V$,
 the behavior of solutions of (\ref{eq:S1}) near $\phi=-\infty$
is equivalent (except of course for the sign of $\phi$) to the behavior of the system near $\phi=\infty$
with the alternative potential $U$. Provided that $U$ is in 
${\cal E}_+^2$, the proceeding analysis on
 $\tilde{\Omega_\epsilon}$ can be
applied (for a suitable choice of $\epsilon$).  In what follows,
 we   shall denote the set of all potentials which are (class k) Well Behaved at both infinity and negative infinity as ${\cal E}^k$. We  denote the exponential order of ${\cal E}^k$ functions at $\infty$ and $-\infty$ by $N_\infty$ and $N_{-\infty}$ respectively.

Note that if we define the set $\Omega (x_0)$ as the region of phase space for which $x<x_0$, then it follows from the fact that $x$ is monotonic increasing that this set is equal to the union of its past orbits.  The procedure outlined above for defining a coordinate system at $-\infty$ allows us to embed  $\Omega (x_0)$ in a smooth compact three dimensional manifold $\Sigma (x_0)$ such that the vector field defined by (\ref{eq:S4}) may be smoothly extended  onto $\Sigma (x_0)$. In order to do this we simply define a coordinate atlas as follows: Firstly,  we define the interior of $\Sigma$ to be the set \mbox{$\{(x,y,\phi): 0<x<x_0, -1 < y < 1 \}$} where the coordinates are as defined in  (\ref{var1.5}). Clearly this set is bounded in $x$ and $y$. On the open subset of this set for which $\phi > \epsilon^{-1}$ for some sufficiently small $\epsilon$ we define a second coordinate patch $(x,y,z)$ according to (\ref{eq:coords}). An analogous coordinate patch $(x,y,z)$ is defined close to $\phi = - \infty$ according to the procedure outlined in the proceeding paragraph. The construction of $\Sigma (x_0)$ is completed by attaching a boundary which is defined by taking the union of the planes $x=0, x= x_0, y= \pm 1, z=0$ on the respective coordinate patches. 

Clearly  $\Sigma (x_0)$ is compact. Indeed it has the topology of a big rectangular box so it could itself be embedded in $R^3$ (Although it is useful to keep this picture in mind, there is no need to carry it out explicitly.) The vector field (\ref{eq:S4}) can be smoothly extended onto the boundary of $\Sigma (x_0)$ such that  $\Sigma (x_0)$ is the union of its past orbits.  

$\Omega (x_0)$ is a 2-dimensional hypersurface embedded in $\Sigma (x_0)$. 
The important feature to note is that $\Omega (x_0)$ approaches the
  (non-physical) boundary
 precisely along the intersection of the plane $x=0$ with the planes $y=\pm 1$
  and $z=0$. We shall term this set, which has the topology of a rectangle, and is independent of $x_0$, the unphysical boundary of $\Omega$, or $\partial\Omega$. Note, in particular, that it contains each of the equilibrium points $p_+, p_-$ and $p_N$ at $\phi = \pm\infty$.  

Keeping all this in mind it is strait-forward to prove the following global singularity theorem.
\begin{theorem}\label{th:flatrwsing}
Let $V$ be a ${\cal E}^2$ potential of exponential order $N^2_{\pm\infty}<6$,   
then there exists a class of FRW cosmologies, containing almost all solutions of (\ref{eq:S4}), which possess an initial space-time singularity. 
Furthermore, these cosmologies are in continuous
 1-1 correspondence with the exactly integrable set of massless scalar field cosmologies. This correspondance is realised asymptotically according to the  the following approximation
\begin{eqnarray}
K&=&t^{-1}+O(\epsilon_V^{\pm}(t))\nonumber\\
\phi&=&\pm\sqrt{\mfrac{2}{3}}\ln\frac{t}{c}+O(t\epsilon_V^{\pm}(t))
\label{eq:stiff}\\
\dot{\phi}&=&\pm\sqrt{\mfrac{2}{3}}t^{-1} + O(\epsilon^{\pm}_V(t))\nonumber
\end{eqnarray}
where
\begin{eqnarray}
\epsilon_V^{\pm}(t)&=&tV(\pm\sqrt{\mfrac{2}{3}}\ln t)\label{eq:err}
\end{eqnarray} 
\end{theorem}
(Sketch) Proof:\\

Clearly, it is sufficient to show that almost all solutions are asymptotic in the past to the equilibrium point $p_-$ (at either $\infty$ or $-\infty$). 
Also, since $x$ is monotonic, it is sufficient to consider solutions on $\Omega (x_0) \subset \Sigma (x_0)$ where $x_0$ is arbitrary. Since $\Sigma (x_0)$ is compact and contains its past orbits, each of its points $p$ must contain an $\alpha$-limit set, $\alpha (p)$,  which is itself a closed union of orbits. In particular, for points on the physical space  $\Omega (x_0)$, Theorem~\ref{th:unbound} implies that $\alpha_p$ must almost always contain at least one point on the sets $z=0$ ($\phi = \pm\infty$). But, from the discussion proceeding the theorem, any point  $z= 0$ which is a limit-point of a physical trajectory must  be part of the unphysical boundary $\partial\Omega $ and must therefore have $x=0$. Since $x$ is monotonic increasing, the entire of $\alpha (p)$ must be contained in the plane $x=0$, or more precisely, on the 1-dimensional ``rectangle'' $\partial \Omega $.  It can easily be shown that the only conceivable generic limit sets on this rectangle are the two equilibrium points corresponding to $p_-$ at
 $\pm\infty$ respectively (By Theorem \ref{th:pN} we know that $p_N$ is not a generic source, neither can the rectangle be a cycle since the sense of orbits reverses at $P_N$). 

This completes the proof. $\Box$ \\   

\section{Discussion} 

 Perhaps the most interesting aspect of Theorem~\ref{th:flatrwsing} is the indication 
it provides that scalar field cosmologies have a simple and quite regular 
asymptotic structure, and that this structure is independent of the exact 
details of the potential. This  allows one to impose initial conditions and 
physically motivated  boundary conditions in an 
unambiguous manner without the need to make an arbitrary choice 
of initial spacelike hypersurface. Rather, one takes the boundary as being the singularity itself and assigns boundary conditions according to the unique asymptotic state of the solution. This corresponds to choosing a particular massless scalar field solution which, in the case of the very simple FRW models above, is equivalent to choosing a parameter value $c$ (together with a choice of sign). 

For example, if one makes the  assumption that all values of $c$ are {\it a priori} equally likely, then it is possible to unambiguously assess the probability of inflation in a particular model by simply comparing the size of the line segments in $c$-space which inflate and don't inflate, respectively. Alternatively, one could carry out quantum gravitational calculations on the massless scalar field cosmologies to obtain a probability distribution $P(c)$ on the set of initial conditions. In view of the technical difficulties associated with quantising scalar field cosmologies with non-zero potentials, whose classical solutions are not generally integrable, the appeal of the above procedure is self evident.

 FRW space-time does not provide a particularly realistic model of the gravitational field close to a singularity due to the tendency of gravitational perturbations, in particular shear anisotropy, to diverge  as the singularity 
is approached. The significance of Theorem~\ref{th:flatrwsing} rests with the relative ease with which it may be generalised to more general space-time
 behavior (and more complicated matter fields). Investigations carried out so far have provided a fairly strong indication that the dynamical insignificance of $V$ close to the singularity is indeed a characteristic feature of scalar field cosmologies in general.  A direct analogue of Theorem \ref{th:flatrwsing} has been proven for Bianchi type I space-time  and a weaker result has been demonstrated for Class A spatially homogeneous space-times, which include Bianchi type VII and IX - the highest dimensional spatially homogeneous cosmologies; ie, it can be shown that almost all solutions can be approximated by $V=0$ solutions asymptotically in the past, under the condition that a bound is placed on positive spatial curvature at some time $t_0$ \cite{thesis}. 

 In fact, the presence of positive spatial curvature introduces significant additional structure into solution space, due to the fact that the time derivative of the expansion $K$ need no longer be monotonic  (as is well known from standard cosmology). For scalar fields this leads to the existence  of singularity free ``bouncing'' solutions for which the spatial volume element  has a non-zero minimum, even in relatively simple FRW models \cite{hawbounce}. For the case of the massive scalar field ($V =1/2m^2\phi^2$)  these solutions have been studied  by Page \cite{page} and, more recently, by a number of authors \cite{khal,shelcorn} who have shown that they comprise a rather complex fractal subset of solution space. Although intriuging, this structure has measure zero in solution space and is therefore non-generic. In models for which $V$ has a non-zero minimum, on the other hand, it appears that bouncing solutions are generic, not only in FRW models but in the more general Bianchi Type IX solution space. The possible physical significance of such solutions is yet to be looked at in detail. Nevertheless, as is the case with no-hair theorems for inflation \cite{wald1,jenstein}, it seems possible to eliminate cosmological bounces by placing a bound on negative curvature. This is suggestive of a close relationship between classes of past asymptotic structure  and  the occurrence of inflation. The past asymptotic structure of the non-bouncing solutions appears to conform with that of negative and zero curvature models, in that neither the curvature, nor the scalar field potential are dynamically significant asymptotically in the past.           

Given that the long wavelength solutions of scalar field cosmologies asymptote towards the solutions of the  massless scalar field ($V\equiv0$) it is significant that such a field is equivalent to a perface fluid with equation of state $\rho =p$; the so called stiff fluid. This extreme case among classical linear-equation-of-state perfect fluids differs from other such models in that its initial singularity is believed to have a velocity dominated semi-Kasner like structure \cite{eard} rather than the more characteristic, and much more complex, oscillatory Belinskii Lifshitz Khalatnikov  singularity \cite{belip1.5,com2}. It is worth noting that the semi-Kasner like singularity structure admits the long wavelength approximation close to the singularity wheras the BKL singularity does not. In both cases however,  the procedure for constructing the generic singularity is essentially to take the general spatially homogeneous solutions (with respect to an orthonormal frame, say) and make the parameters dependent on spatial coordinates. Thus, as far as we know, the asymptotic behavior of general cosmological perfect fluid solutions is very closely related to, and is indeed characterised by, the solution space of spatially homogeneous space-times.          

More work is required before we can state with confidence that this is also true of scalar field cosmologies but it seems probable  that it will  at least
 be 
 true of a generic family of solutions, provided one is careful to insist 
that the scalar field has non-negative energy (ie; that the weak energy condition is satisfied). This  condition may be justified on heuristic grounds, and has the effect of imposing significant 
constraints on the degree of inhomogeneity of the scalar field. The author would further speculate, that the weak energy condition is also likely to be a necessary condition in  ensuring that significant proportion of solutions undergo inflation in accordance with the already stated view  that there exists a close relationship between singularity structure and inflation.

\begin{center}
{\large \bf Acknowledgments}
\end{center} 
The author would particularly like to thank Peter Szekeres for  helpful comments and suggestions on the content and layout of this article and Donald Oats for many enlightening  discussions on dynamical systems theory. This research was supported in part by an Australian Post Graduate Research
 Award.

\end{document}